\documentclass[12pt]{article}

\usepackage{graphicx,epsfig,epsf,amssymb}

\usepackage{scicite}

\usepackage{times}

\def\gsim{\;\lower4pt\hbox{${\buildrel\displaystyle >\over\sim}$}\;}
\def\lsim{\;\lower4pt\hbox{${\buildrel\displaystyle <\over\sim}$}\;}
\def\grls{\;\lower4pt\hbox{${\buildrel\displaystyle >\over <}$}\;}

\newenvironment{sciabstract}{%
\begin{quote} \bf}
{\end{quote}}

\newcounter{lastnote}
\newenvironment{scilastnote}{%
\setcounter{lastnote}{\value{enumiv}}%
\addtocounter{lastnote}{+1}%
\begin{list}%
{\arabic{lastnote}.}
{\setlength{\leftmargin}{.22in}}
{\setlength{\labelsep}{.5em}}}
{\end{list}}

\title{\bf Anisotropy and corotation \\
           of Galactic cosmic rays }

\author{
M.~Amenomori $^{1}$,
S.~Ayabe $^{2}$,
X.J.~Bi $^{3}$,
D.~Chen $^{4}$,
S.W.~Cui $^{5}$,\\
Danzengluobu $^{6}$,
L.K.~Ding $^{3}$,
X.H.~Ding $^{6}$,
C.F.~Feng $^{7}$,
Zhaoyang~Feng $^{3}$,\\
Z.Y.~Feng $^{8}$,
X.Y.~Gao $^{9}$,
Q.X.~Geng $^{9}$,
H.W.~Guo $^{6}$,
H.H.~He $^{3}$,\\
M.~He $^{7}$,
K.~Hibino $^{10}$,
N.~Hotta $^{11}$,
Haibing~Hu $^{6}$,
H.B.~Hu $^{3}$,\\
J.~Huang $^{12}$,
Q.~Huang $^{8}$,
H.Y.~Jia $^{8}$,
F.~Kajino $^{13}$,
K.~Kasahara $^{14}$,\\
Y.~Katayose $^{4}$,
C.~Kato $^{15}$,
K.~Kawata $^{12}$,
Labaciren $^{6}$,
G.M.~Le $^{16}$,\\
A.F.~Li $^{7}$,
J.Y.~Li $^{7}$,
Y.-Q.~Lou $^{17}$,
H.~Lu $^{3}$,
S.L.~Lu $^{3}$,
X.R.~Meng $^{6}$,\\
K.~Mizutani $^{2,18}$,
J.~Mu $^{9}$,
K.~Munakata $^{15}$,
A.~Nagai $^{19}$,
H.~Nanjo $^{1}$,\\
M.~Nishizawa $^{20}$,
M.~Ohnishi $^{12}$,
I.~Ohta $^{21}$,
H.~Onuma $^{2}$,
T.~Ouchi $^{10}$,\\
S.~Ozawa $^{12}$,
J.R.~Ren $^{3}$,
T.~Saito $^{22}$,
T.Y.~Saito $^{12}$,
M.~Sakata $^{13}$,\\
T.K.~Sako $^{12}$,
T.~Sasaki $^{10}$,
M.~Shibata $^{4}$,
A.~Shiomi $^{12}$,
T.~Shirai $^{10}$,\\
H.~Sugimoto $^{23}$,
M.~Takita $^{12}$,
Y.H.~Tan $^{3}$,
N.~Tateyama $^{10}$,
S.~Torii $^{18}$,\\
H.~Tsuchiya $^{24}$,
S.~Udo $^{12}$,
B.~Wang $^{9}$,
H.~Wang $^{3}$,
X.~Wang $^{12}$,\\
Y.G.~Wang $^{7}$,
H.R.~Wu $^{3}$,
L.~Xue $^{7}$,
Y.~Yamamoto $^{13}$,
C.T.~Yan $^{12}$,\\
X.C.~Yang $^{9}$,
S.~Yasue $^{25}$,
Z.H.~Ye $^{16}$,
G.C.~Yu $^{8}$,
A.F.~Yuan $^{6}$,\\
T.~Yuda $^{10}$,
H.M.~Zhang $^{3}$,
J.L.~Zhang $^{3}$,
N.J.~Zhang $^{7}$,
X.Y.~Zhang $^{7}$,\\
Y.~Zhang $^{3}$,
Yi~Zhang $^{3\ast}$,
Zhaxisangzhu $^{6}$,
and X.X.~Zhou $^{8}$ \\
(The Tibet AS$\gamma$ Collaboration) \\
\normalsize{$^\ast$To whom correspondence should
be addressed; E-mail: zhangyi@mail.ihep.ac.cn}  }

\date{}

\begin{document}
\baselineskip24pt

\maketitle

{\footnotesize
\begin{enumerate}
\item{ Department of Physics, Hirosaki University, Hirosaki 036-8561, Japan}
\item{ Department of Physics, Saitama University, Saitama 338-8570, Japan }
\item{ Key Laboratory of Particle Astrophysics, Institute of High Energy
Physics, Chinese Academy of Sciences, Beijing 100049, China }
\item{ Faculty of Engineering, Yokohama National University,
Yokohama 240-8501, Japan }
\item{ Department of Physics, Hebei Normal University,
Shijiazhuang 050016 , China }
\item{ Department of Mathematics and Physics,
Tibet University, Lhasa 850000, China }
\item{ Department of Physics, Shandong University, Jinan 250100, China }
\item{ Institute of Modern Physics, South West Jiaotong University,
Chengdu 610031, China }
\item{ Department of Physics, Yunnan University, Kunming 650091, China }
\item{ Faculty of Engineering, Kanagawa University, Yokohama 221-8686, Japan}
\item{ Faculty of Education, Utsunomiya University, Utsunomiya 321-8505, Japan}
\item{ Institute for Cosmic Ray Research,
University of Tokyo, Kashiwa 277-8582, Japan }
\item{ Department of Physics, Konan University, Kobe 658-8501, Japan}
\item{ Faculty of Systems Engineering, Shibaura
Institute of Technology, Saitama 337-8570, Japan}
\item{ Department of Physics, Shinshu University, Matsumoto 390-8621, Japan}
\item{ Center of Space Science and Application Research,
Chinese Academy of Sciences, Beijing 100080, China }
\item{ Physics Department and Tsinghua Center for Astrophysics,
Tsinghua University, Beijing 100084, China }
\item{ Advanced Research Institute for Science and Engineering,
Waseda University, Tokyo 169-8555, Japan}
\item{ Advanced Media Network Center, Utsunomiya University,
Utsunomiya 321-8585, Japan}
\item{ National Institute of Informatics, Tokyo 101-8430, Japan}
\item{ Tochigi Study Center, University of the Air, Utsunomiya 321-0943, Japan}
\item{ Tokyo Metropolitan College of
Industrial Technology, Tokyo 116-8523, Japan}
\item{ Shonan Institute of Technology, Fujisawa 251-8511, Japan}
\item{ RIKEN, Wako 351-0198, Japan}
\item{ School of General Education, Shinshu University,
Matsumoto 390-8621, Japan}
\end{enumerate}
}

\begin{sciabstract}
The intensity of Galactic cosmic rays is nearly isotropic
due to the influence of magnetic fields in the Milky Way.
Here we present two-dimensional high-precision anisotropy measurement
for energies from a few to several hundred TeV using the huge data
sample of the Tibet Air Shower Arrays. Besides revealing finer details
of the known anisotropies, a new component of sidereal time Galactic
cosmic ray anisotropy
is uncovered around the Cygnus region direction. For cosmic-ray
energies up to a few hundred TeV, all components of anisotropies fade
away, showing a corotation of Galactic cosmic rays with the local
Galactic magnetic environment. These results bear broad implications
to cosmic rays, supernovae, magnetic field, heliospheric and Galactic
dynamic environment in a comprehensive manner.
\end{sciabstract}

\section*{Galactic cosmic rays}
The anisotropy of Galactic cosmic rays (GCRs) may result from an
uneven distribution of cosmic ray (CR) sources and the process of
CR propagation in the Milky Way. CRs of energy below $10^{15}$eV
are accelerated by diffusive magnetohydrodynamic (MHD)
shocks~\cite{Voelk2003,Lou1994,Dar2005,YuLou2006} of supernova 
remnants (SNRs) and stellar winds. The discreteness of SNRs could 
lead to a CR anisotropy~\cite{SNR}.
However, GCRs must almost completely lose their original directional
information; their orbits are deflected by the Galactic magnetic field
(GMF) and are randomized by irregular GMF components, having traveled
on average for many millions of years, some also having interacted with
interstellar gas atoms and dusts.
The transport of CRs in a magnetized plasma is governed by four major
processes: convection, drift, anisotropic diffusion and adiabatic energy
change (deceleration or acceleration)~\cite{Krymsky1964,Parker1965}. High-precision
measurement of the CR anisotropy provides a means to explore magnetic
field structures and gains insight for the CR transport
parameters~\cite{Heber2001}.
The long-term high-altitude observation at the Tibet Air Shower Arrays
(referred to as the Tibet AS$\gamma$ experiment) has accumulated tens
of billions of CR events in the multi-TeV energy range, ready for an
unprecedented high-precision measurement of the CR anisotropy as well
as the temporal and energy dependence of the CR anisotropy.

An expected anisotropy is caused by the relative motion between
the observer and the CR plasma, known as the Compton-Getting (CG)
effect~\cite{CG} with CRs arriving more intense from the motion
direction and less intense from the opposite direction. Such a
CR anisotropy, caused by the Earth orbital motion around the
Sun, has indeed been detected~\cite{Nat,Amenomori2004}.
Data assembled to 1930s~\cite{CG} were consistent with a scenario
that the CR plasma stays at rest in an inertial frame of reference
attached to the Galactic center. If this were true, the Galactic
rotation in the solar neighborhood might then be measurable.
Nevertheless, such CR anisotropy due to the solar system rotation
around the Galactic center at a speed of $\sim 220\hbox{ km s}^{-1}$
remained inconclusive over seven decades. Now our high-precision
two-dimensional (2D) measurement gives a strong evidence to exclude
the CR anisotropy of this origin and thus show a corotation of GCRs
with the local GMF environment.

Historically, the GCR anisotropy~\cite{Nagashima1976,Nagashima1998} 
has been measured as the sidereal time variation at the spinning 
Earth using both underground $\mu$ detectors and ground-based air 
shower arrays~\cite{KMunakata,MARCO,EAS-TOP,KASCADE,SK-I,GA-Tibet}.
Located at different geographical latitudes and operating in 
different years with various threshold energies, each individual 
experiment could only measure the CR modulation profile along the 
R.A. direction which was usually fitted by first few harmonics. 
Instead of using sine or cosine harmonics, one may adopt two 
Gaussian functions with Declination dependent parameters (mean, 
width and amplitude) to fit the so-called ``tail-in" and 
``loss-cone" features, and a tentative 2D anisotropic picture was 
obtained~\cite{Hall1998,Hall1999} by simultaneously fitting 
different experimental data. Here, the CR deficiency was thought 
to be associated with a magnetic cone-like structure and thus the 
name `loss-cone', while the CR enhancement is roughly in the 
direction of the heliospheric magnetotail and is thus referred 
to as `tail-in' enhancement~\cite{Nagashima1976,Nagashima1998}.
However, the accurate spatial and energy dependence of CR 
anisotropy could not be given~\cite{Duldig}, and some subtle 
features remain hard to reveal, since the CR anisotropy is more 
complex and cannot be properly described by two Gaussians. The 
Tibet AS-$\gamma$ experiment alone can achieve 2D measurement 
in various energy ranges and to provide details of the 2D CR 
anisotropy.

\section*{Tibet Air Shower Array Experiment}

The Tibet Air Shower Array experiment has been conducted at Yangbajing
(90.522 E, 30.102 N; 4300 m above the sea level) in Tibet, China since
1990. The Tibet I array~\cite{TB1992}, consisting of 49 scintillation
counters and forming a 7$\times$7 matrix of 15 m span, was expanded to
become the Tibet II array with an area of 36,900 m$^2$ by increasing
the number of counters in 1994.
In 1996, part of Tibet II with an area of 5175 m$^2$ was upgraded to a
high-density (HD) array with a 7.5 m span~\cite{TB2001b}. To increase
the event rate, the HD array was enlarged in 1999 to cover the central
part of Tibet II as Tibet III array~\cite{TB2001c,TB1993,TB2003}. The area of Tibet
III array has reached 22,050 m$^2$.
The trigger rates are $\sim$ 105Hz and $\sim$ 680Hz for the
Tibet HD and III arrays, respectively. The data were acquired
by running the HD array for 555.9 live days from 1997.2
to 1999.9
and the Tibet III array for 1318.9 live days from 1999.11
to 2005.10.
GCR events are selected, if any four-fold coincidence occurs
in the counters with each recording more than 0.8 particles in
charge, the air shower core position is located in the array
and the zenith angle of arrival direction is  $\lsim 40^\circ$.
With all those criteria, both Tibet HD and III arrays have
the modal energy of 3 TeV and a moderate energy resolution;
the $\sim 0.9^{\circ}$ angular resolution estimated from
Monte Carlo simulations~\cite{TB1990,Kasahara}
was verified by the Moon shadow measurement~\cite{TB2001c,TB1993,TB2003}.
In total, $\sim$ 37 billion CR events are used in our data analysis.

\section*{Data analysis and results}

With such a large data sample, we conduct a 2D measurement to reveal 
detailed structural information of the large-scale GCR anisotropy beyond 
the simple R.A. profiles. For each short time step (e.g. 2 minutes), the 
relative CR intensity at points in each zenith angle belt can be compared 
and this comparison can be extended step by step to all points in the 
surveyed sky [see Refs. \cite{All-Sky} for details of 
data analysis]. Lacking the absolute detector efficiency calibration in 
the Dec direction, absolute CR intensities along different Dec directions 
cannot be compared. Thus, the average intensity in each narrow Dec belt 
is normalized to unity. Our analysis procedure would give a correct 2D 
anisotropy if there is no variation in the average CR intensity for 
different Dec.
We systematically examined the CR anisotropy in four different time 
frames, namely solar time for solar modulation, sidereal time for 
Galactic modulation, anti-sidereal time and extended-sidereal time; 
and systematic variations are found to be unimportant.

To study the temporal variation of CR modulation, we divide the data
sample into two subsets. The first subset is from 1997.2 to 2001.10
covering the 23rd solar maximum (a period of a few
years when solar magnetic activity are the strongest)
while the second subset is from 2001.12 to 2005.11
approaching the solar minimum (a period of a few years when solar
magnetic activity become minimal). Comparing the sidereal time
plots for these two intervals (Fig. 1) shows that the
CR anisotropy is fairly stable and insensitive to solar activity.
The `tail-in' and `loss-cone' anisotropy components~\cite{Nagashima1976,Nagashima1998},
extracted earlier from a combination of the underground $\mu$ telescope
data analyses~\cite{Hall1998,Hall1999}, are seen in our 2D plots in much
finer details and with a high significance (Fig. 1).
Our new high-precision measurement thus provides
constraints on physical interpretations of these features.

Spreading across $\sim 280^{\circ}$ to $\sim 360^{\circ}$ in R.A., a
new excess component with a $\sim 0.1$\% increase of the CR intensity
peaked at Dec $\sim 38^\circ$ N and R.A. $\sim 309^\circ$ in the
Cygnus region is detected at a high significance level of
$13.3\sigma$ with a $5^\circ$ pixel radius (Fig. 1d).
The Cygnus region, where complex features are revealed in broad
wavelength bands of radio, infrared, X-ray and gamma-rays, is
rich of candidate GCR sources. Recently, the first unidentified
TeV gamma-ray source was discovered here by HEGRA~\cite{HEGRA}.
This region, as observed by EGRET~\cite{EGRET}, appears to be the
brightest source of diffuse GeV gamma-rays in the northern sky and
contributes significantly to the diffuse TeV gamma-ray emission in the
Galactic plane as observed by Milagro~\cite{MILAGRO} which rejects 90\%
of CR background while still retains $\sim$45\% of gamma-rays. Such
gamma-rays originate from the interaction of CRs with gas and dusts.
Using more stringent event selection criteria~\cite{All-Sky}, a deeper
view of Cygnus region with a $0.9^\circ$ pixel radius shows that the
large-scale excess consists of a few spatially separated excesses of
smaller scales superposed onto a large-scale anisotropy (Fig. 1e);
these small-scale ($\sim 2^\circ$) excess favors the interpretation that
the extended gamma-ray emission from the Cygnus region contributes
significantly to the overall excess in the region~\cite{Kieda}.
As our experiment cannot yet distinguish gamma-rays from the charged
CR background, we cannot tell how much of this excess is to be
attributed to gamma-rays and how much, if any, is associated with
charged CRs~\cite{Berg-Orion}. Such a determination requires upgrading
the Tibet arrays for CR and gamma-ray discrimination.

The solar time CR modulation was also stable (Fig. 2). We
found that including events with fewer than 8 detector coincidences
(lower energy events) resulted in much larger modulation amplitudes
than those obtained when these events were excluded (higher energy
events). To avoid this, high multiplicity events with coincident
detector numbers $\geq 8$ were adopted (Fig. 2). The
observed dipole anisotropy agrees very well with the expected CG
effect due to the Earth orbital motion around the Sun. Thus,
heliospheric magnetic field and solar activity does not influence
the multi-TeV CR anisotropy.

Because of the stable nature of the sidereal time modulation, data
from different years were combined to examine the energy dependence of
CR anisotropy. Fig. 3 shows the variation of anisotropy
for five groups of events according to their different primary
energies. For primary energies below 12 TeV, the anisotropies show
little dependence on energy while above this energy, anisotropies
fade away, consistent with a CR isotropy of KASCADE~\cite{KASCADE}
in the energy range of $0.7-6$PeV. Contrary to the earlier
suggestion~\cite{Nagashima1998}, the `tail-in' component remains still visible
above 50 TeV in smaller regions. Since the multi-TeV GCRs, whose
gyro-radii are hundreds or thousands of AUs, are not affected by
the heliospheric magnetic field, it is clear that the GMF must be
responsible for both `tail-in' and `loss-cone' modulations.

As a result of a diminishing GCR anisotropy at high energies,
we can test the CG anisotropy caused by the orbital motion of
the solar system in our Galaxy, which would peak at
($\alpha=315^\circ,~\delta= 49^\circ$) and minimize at
($\alpha=135^\circ,~\delta=-49^\circ$) with an amplitude of 0.35\%.
This would be a salient signal in a real 2D measurement. However, as
explained earlier, the modulation along the Dec direction is partly lost.
After applying the normalization procedure along each Dec belt, the
expected CG anisotropy is distorted and apparently peaks at ($\alpha=
315^\circ,~\delta=0^\circ$) and forms a trough around ($\alpha=135^\circ,
~\delta=0^\circ$) with a smaller amplitude of $\sim$0.23\% (Fig. 4).
To avoid any contamination from the non-vanishing `tail-in' and `loss-cone'
anisotropies~\cite{Nagashima1976,Nagashima1998} when the primary energy is of $\sim 300$ TeV, the
upper half of the surveyed CR intensity map (with Dec $>25^\circ$) is used
to compare with the predicted Galactic CG effect of an amplitude
$\sim$0.16\%. The fitted anisotropy amplitude is $0.03$\%$\pm0.03$\%,
consistent with an isotropic CR intensity. Therefore our observation
exclude the existence of the Galactic CG effect with a high-degree
confidence, assuming the absence of other cancelling effects. The null
result of the Galactic CG effect implies that GCRs corotate with the
local GMF environment.

\section*{Discussion}

The observation of GCR anisotropy and diffuse gamma-ray emission
plays an important role in probing sources and propagation of CRs.
The detection of the new large-scale GCR anisotropy component
and the indication of extended gamma-ray emission
from the same mysterious Cygnus region allow us to
connect the GCR acceleration site and propagation.
A precision spectral and morphological determination of the
extended gamma-ray emission would be our next pursuit.
The existence of large-scale GCR
anisotropies up to a few tens of TeV
indicates that they are not related to the heliospheric magnetic
field. It is conceivable that GMF has large-scale structures in
the heliospheric neighborhood.

As in many spiral galaxies, our Milky Way
has large-scale differential rotations in stellar and magnetized gas
disks with a GMF of a few micro Gauss. The GMF, GCRs and thermal gas
have similar energy densities of $\sim 1\hbox{eV cm}^{-3}$ and
interact with each other dynamically.
The corotation of the GCR plasma with the local GMF
environment around the Galactic center is enforced
by the Lorentz force as GCRs randomly scatter and
drift in irregular GMF components~\cite{ParkerEN}.
As the Galactic disk rotates differentially, the important
inference is that the bulk GCR plasma within and above the
Galactic disk must also rotate differentially.
The GCR corotation evidence provides an important empirical
basis for the study of galactic MHD processes, such as
modeling synchrotron emission diagnostics for large-scale
spiral structures of MHD density waves~\cite{FanLou1996,LouFan2003,LouBai2006}.
%
\bibliography{scibib}
\bibliographystyle{Science}

\begin{scilastnote}
\item

The collaborative experiment of the Tibet Air
Shower Arrays has been under the auspices of
the Ministry of Science and Technology of
China and the Ministry of Foreign Affairs
of Japan. 
This work was supported in part 
by Grants-in-Aid for Scientific Research on Priority Areas (712) 
(MEXT), by Scientific Research (JSPS) in Japan, by the National 
Natural Science Foundation of China 
and by the Chinese Academy of Sciences. The authors thank 
J. K\'ota for reading the  manuscript and critical comments.
\end{scilastnote}

\begin{figure}
\begin{center}
  \includegraphics[height=16cm, width=16cm]{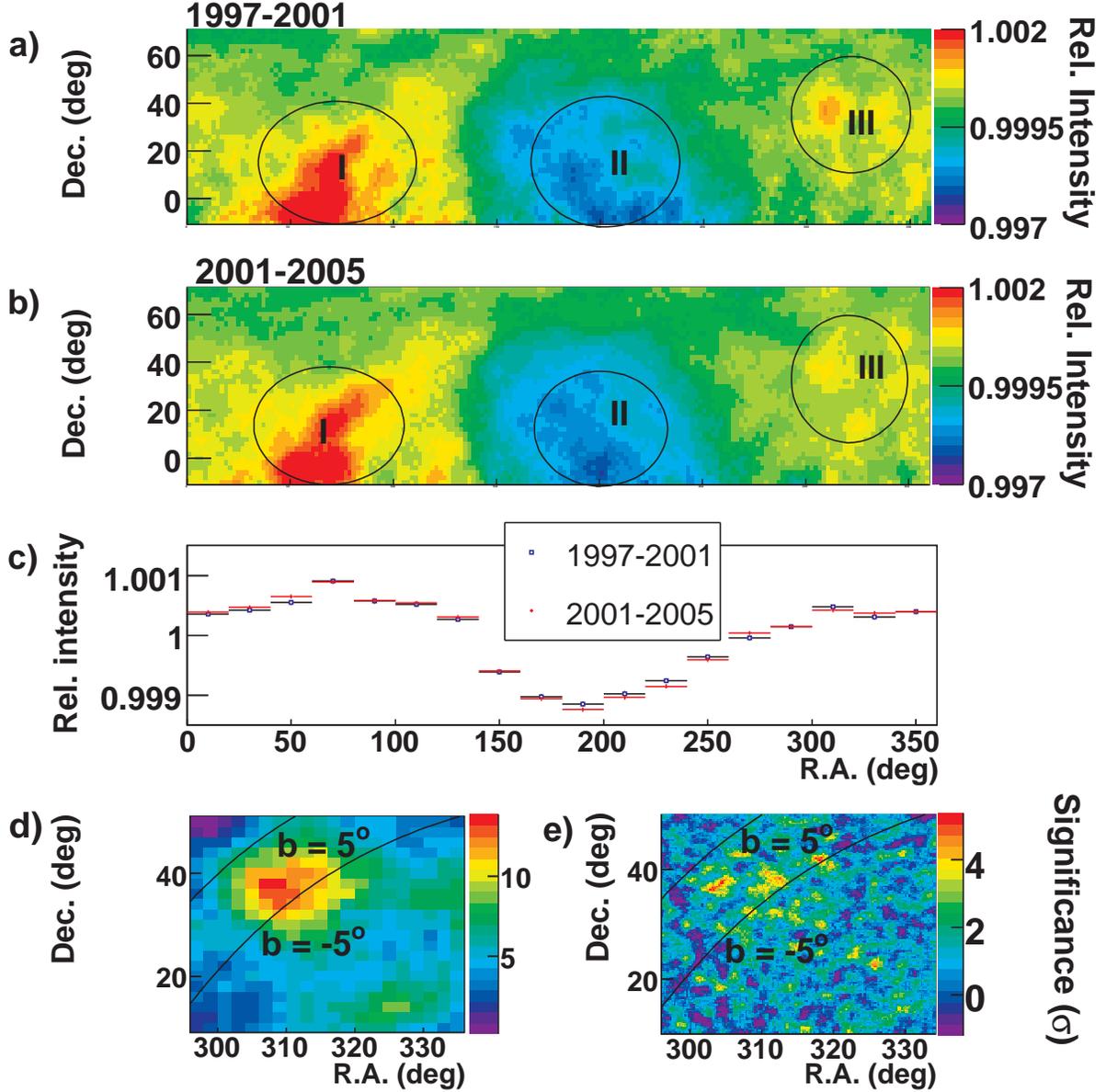}
\end{center}
\caption {\footnotesize Celestial CR intensity map$^{\dagger}$
for Tibet AS$\gamma$ data taken from 1997$-$2001 (a) and
2001$-$2005 (b).
The vertical color bin width is $2.5\times 10^{-4}$
for the relative intensity in both (a) and (b).
The circled regions labeled by I, II and III are the `tail-in',
`loss-cone'~\cite{Nagashima1976,Nagashima1998} and newly found anisotropy
component around the Cygnus region ($\sim 38^\circ$ N Dec and
$\sim 309^\circ$ R.A.), respectively. Panel (c) is the
one-dimensional (1D) projection of the 2D maps in R.A.
for comparison.
Panels (d) and (e) show the significance
maps of Cygnus region (pixels in radius
of $0.9^\circ$ and sampled over a
square grid of side width
$0.25^\circ$ for (e))
for data of 1997$-$2005. The vertical color bin widths are
$0.69\sigma$ and $0.42\sigma$ for the significance in (d)
and (e), respectively.
Two thin curves in both (d) and (e) stand for the Galactic parallel
$b=\pm 5^\circ$. Small-scale anisotropies (e) superposed onto the
large-scale anisotropy
hint at the extended gamma-ray emission.
$^{\dagger}$Unless otherwise stated, images in Figs 1$-$4 are
presented using pixels in radius of $5^\circ$ and sampled over a
square grid of side width $2^\circ$; the modal energy is 3 TeV.  
}
\end{figure}

\begin{figure}
\begin{center}
\includegraphics[height=16cm, width=16cm]{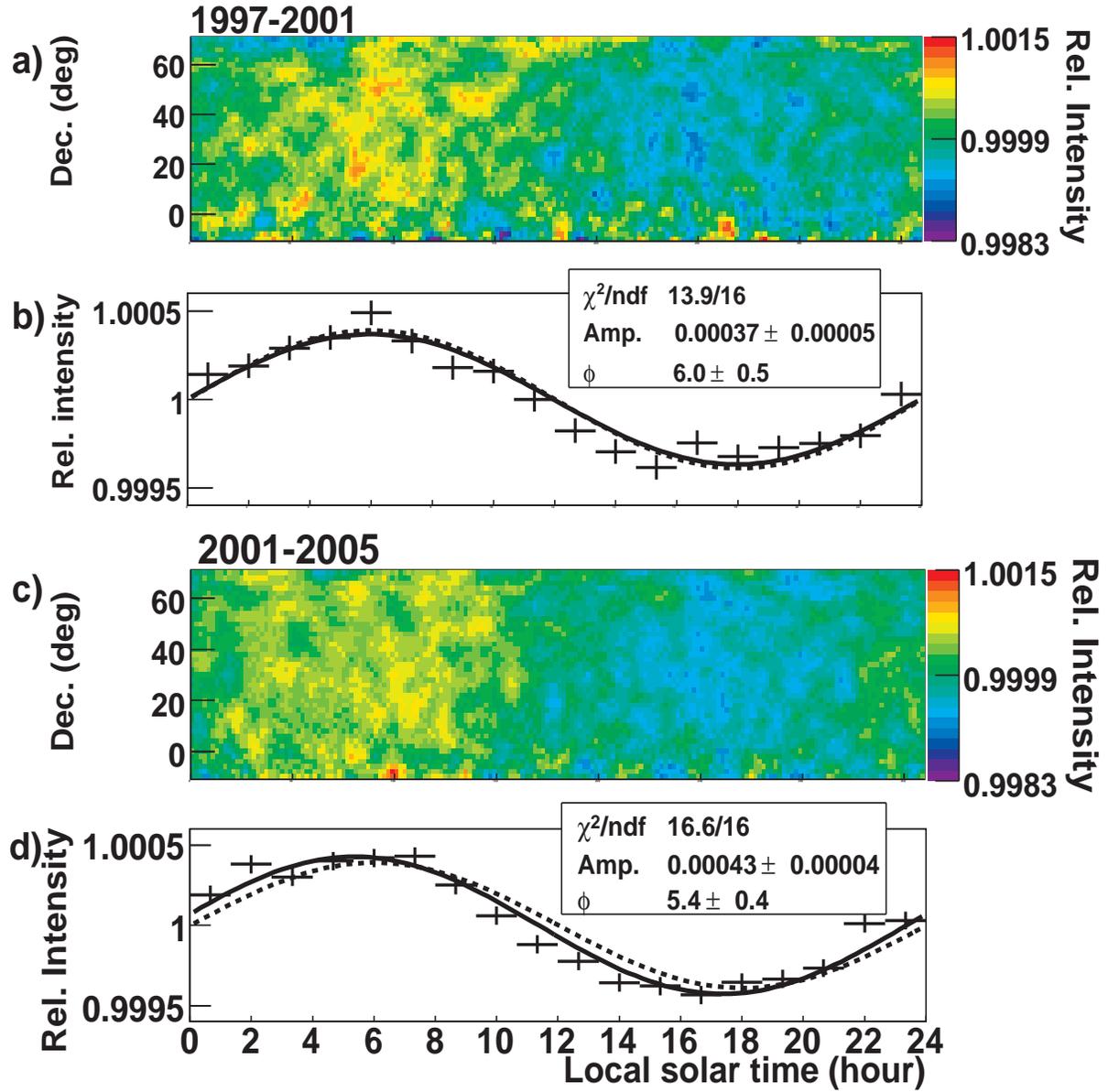}
\end{center}
\caption {\footnotesize
Local solar time CR intensity map
for the Tibet
AS$\gamma$ data taken from 1997$-$2001 (a) and (b) and 2001$-$2005
(c) and (d). Both samples have the modal energy of 10 TeV. The vertical
color bin width is $1.6\times 10^{-4}$ for the relative intensity in
both (a) and (c). In both (b) and (d), the fitting function is in the
form of $\hbox{ Amp}*\cos[2\pi\hbox{(T$-\phi$)}/24]$ where the local
solar time T and $\phi$ are in unit of hour and Amp is the amplitude.
The $\chi^2$ fit involves the number of degree of freedom (ndf) given
by the number of bins minus 2 due to the two fitting parameters Amp and
$\phi$. The 1D plots are the projection of the 2D maps in local solar
time. In the 1D plots the dashed lines are from the expected CG effect,
while the solid lines are the best harmonic fits, which agree very
well with the prediction. The solar time modulation appears quite
stable and insensitive to solar activity. 
}
\end{figure}

\begin{figure}
\begin{center}
\includegraphics[height=16cm, width=16cm]{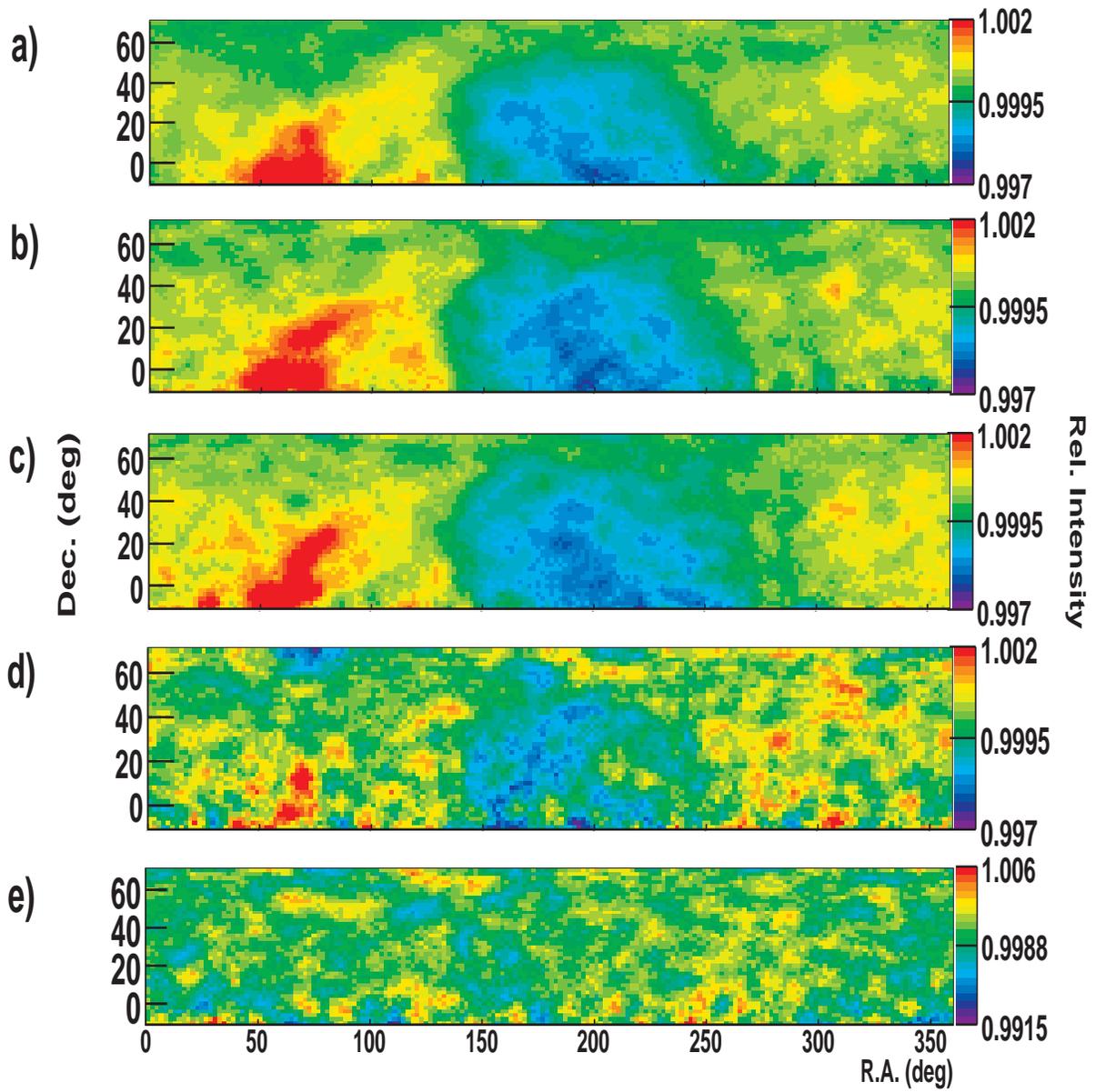}
\end{center}
\caption {\footnotesize
Celestial CR intensity map
for different representative CR energies: (a) 4
TeV; (b) 6.2 TeV; (c) 12 TeV; (d) 50 TeV; (e) 300TeV. Data
were taken during 1997$-$2005. The vertical color
bin width is $2.5\times 10^{-4}$ in (a)$-$(d), while it is
$7.25\times 10^{-4}$ in (e) for different statistics, all
for the relative CR intensity.
}
\end{figure}

\begin{figure}
\begin{center}
  \includegraphics[height=16cm, width=16cm]{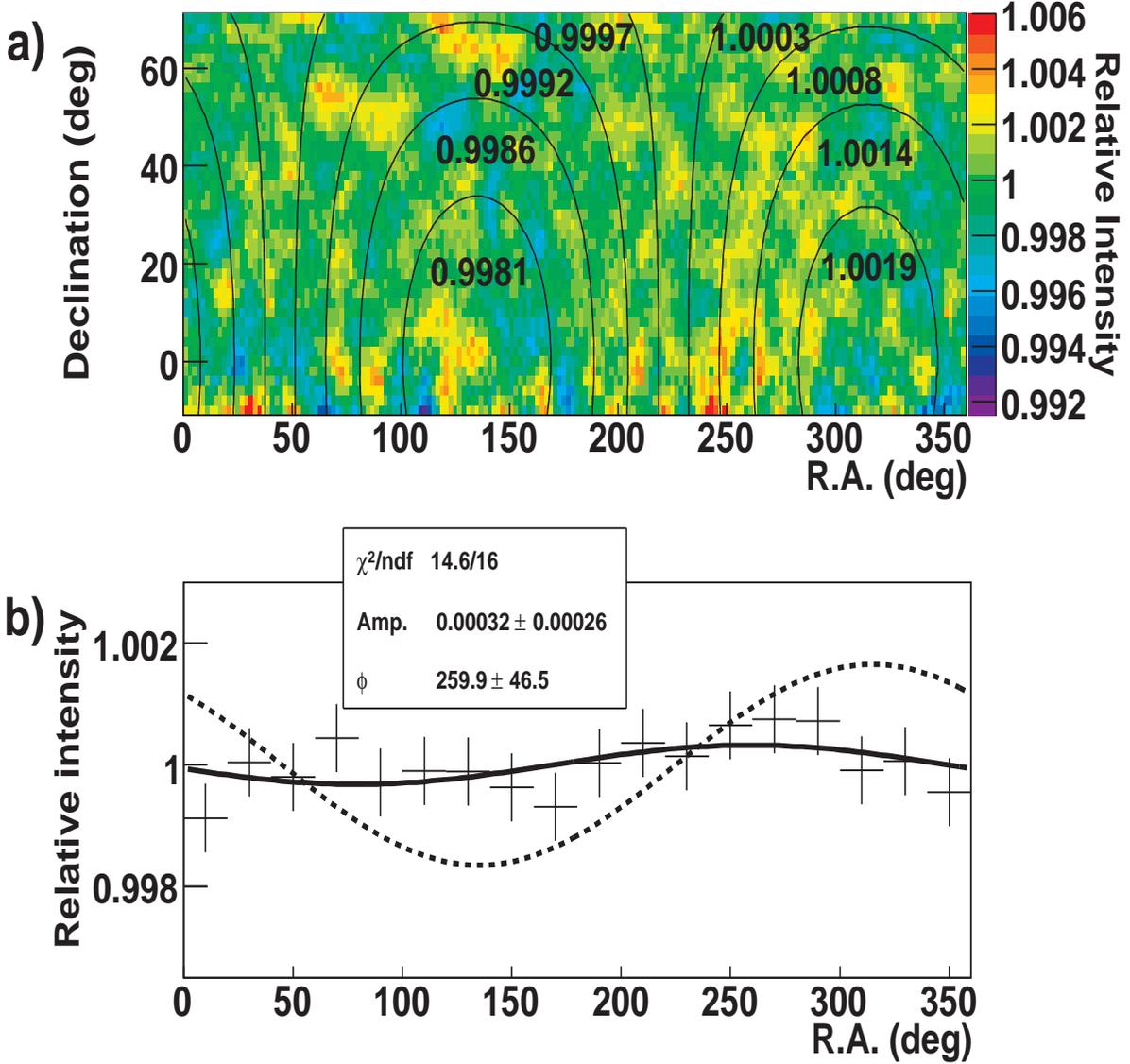}
\end{center}
\caption {\footnotesize
 Celestial or 2D local sidereal time CR
intensity map and its 1D projection in the R.A. direction for
300 TeV CRs of all data. (a) The colored map is the same as
Fig. 3e, while the contours are the ``apparent'' 2D anisotropy
expected from the Galactic CG effect. The vertical color bin
width is $7.25\times 10^{-4}$ for the relative intensity in (a).
The 1D projection is in map (b) for Dec between $25^\circ-70^\circ$,
where the dashed line is the expected Galactic CG response while the
solid line is the best fit to this observation using a first-order
harmonic function. The fitting function is in the form of
$\hbox{ Amp}*\cos(\hbox{R.A.}-\phi)$ where $\phi$ is in degree and
Amp is the amplitude. The $\chi^2$ fit involves the ndf given by
the number of bins minus 2 for the two fitting parameters Amp and
$\phi$. The data shows no Galactic CG effect with a high-level
of confidence. 
}
\end{figure}
\end{document}